\documentclass[figure]{pasj00}
\Received{$\langle$2008 October 2$\rangle$}
\Accepted{$\langle$2008 December 17$\rangle$}
\Published{$\langle$publication date$\rangle$}
\SetRunningHead{Matsui \& Habe}{Dynamical Evolution of a Supermassive Black Hole Binary}

\usepackage{times}

\begin{document}

\title{Effects of Minor Mergers on
Coalescence of a Supermassive Black Hole Binary}
\author{Hidenori \textsc{Matsui}}
\affil{%
   Division of Theoretical Astronomy, 
   National Astoronomical Observatory of Japan, 2--21--1 Osawa,\\
   Mitaka-shi, Tokyo 181--8588}
\email{hidenori.matsui@nao.ac.jp}
\and
\author{Asao \textsc{Habe}}
\affil{Division of Physics,
Graduate School of Science,
Hokkaido University, Sapporo 060, Japan}
\email{habe@astro1.sci.hokudai.ac.jp}

\maketitle

\begin{abstract}
We study the possibility
that minor mergers resolve
the loss cone depletion problem,
which is the difficulty occured in the coalescence process
of the supermassive black hole (SMBH) binary,
by performing numerical simulations
with a highly accurate $N$-body code.
We show that
the minor merger of a dwarf galaxy disturbs
stellar orbits in the galactic central region
of the host galaxy
where the loss cone depletion is already caused
by the SMBH binary.
The disturbed stars
are supplied into the loss cone.
Stars of the dwarf galaxy are also supplied into the loss cone.
The gravitational interactions
between the SMBH binary and these stars
become very effective.
The gravitational interaction decreases
the binding energy of the SMBH binary effectively.
As a result,
the shrink of the separation of the SMBH binary is accelerated.
Our numerical results strongly suggest that
the minor mergers are one of the important processes
to reduce the coalescence time of the SMBH binary
much less than the Hubble time.
\end{abstract}

\section{Introduction}
It is well known that
the merging of galaxies
with central supermassive black holes (SMBHs)
is an important process
for growth of the SMBH mass.
Recent studies,
however,
have shown the possibility that
the SMBHs cannot coalesce
within the Hubble time
in the merger remnant
\citep{beg80,mak04}.

\citet{beg80} have examined the merging process
of galaxies with central SMBHs with masses of $10^8 M_{\odot}$.
They estimate the timescale of coalescence of SMBHs
in the merger remnant
and show that
the timescale is more than the Hubble time.
They describe the reason for this as follows:
SMBHs sink into the center
in the merger remnant
because of the dynamical friction
from field stars.
During this process,
the field stars near the center are scattered
by SMBHs
and the number of the field stars decreases.
The loss cone depletion occurs
because of the scattering.
In this case,
dynamical friction force on the SMBHs becomes very weak.
Then,
it is difficult for the SMBH binary
to shrink to a small distance
enough to emit significant gravitational waves
in the final coalescence stage of SMBHs.

\citet{mak04} have studied the dynamical evolution of a SMBH binary,
that each mass of the SMBH is $10^{8} M_{\odot}$,
in the stellar system
by performing high-resolution $N$-body simulations.
Their results have shown that
the hardening timescale of the binary strongly depends on
the relaxation time of the host galaxy
as predicted by \citet{beg80}.
These results have confirmed the prediction that
the SMBH binary
cannot coalesce
within the Hubble time
by only gravitational interactions
between SMBHs and field stars of the host galaxy
since the relaxation time in a galactic stellar system
is larger than the Hubble time.
This difficulty in the coalescence process of SMBHs
is called the "loss cone depletion problem".

To resolve the loss cone depletion problem,
several ideas to accelerate the orbital decay of the binary are proposed.
Gaseous torque in a massive gas disk
is proposed in the wet merger cases
\citep{esc04,esc05,dot06,dot07,hay08}.
In the dry merger cases which are observed in neaby galaxies \citep{whi08},
the effect of a galactic triaxial potential \citep{ber06},
large mass ratio between a SMBH and Intermediate-Mass BH
\citep{mat07},
and a triple SMBH system \citep{iwa06}
have been proposed.
In the triple SMBH system,
two SMBHs coalesce through the three body instability and the Kozai mechanism
and the coalescence possibility is roughly $50$ \% \citep{iwa06}.
\citet{per07} and \citet{per08} have proposed the role of massive perturbers.
In their analytical studies,
it has been shown the possibility that
the massive perturbers of giant molecular clouds or molecular gas clumps
accelerate the relaxation of stars
in the galactic central region
and, as a result,
trigger the rapid coalescence of the SMBH binary.
They pointed out the importance of three body interactions
between the SMBH binary and stars.
These ideas have possibility
to lead to the coalescence of two SMBHs
within the Hubble time,
if some suitable conditions are realized.

In this paper,
in order to resolve the loss cone depletion problem,
we propose a new scenario
that a minor merger triggers the rapid shrink of the SMBH binary.
Similar idea was studied by \citet{per07} and \citet{per08}
Our scenario is as follows:
If a dwarf galaxy is compact enough,
it can come close to the galactic center,
and then,
stellar orbits of the host galaxy are highly disturbed
by the dwarf galaxy.
In this case,
many stars will be supplied into the loss cone.
Moreover,
if the dwarf galaxy is not destroyed
before it closes enough to the central region,
stars of the dwarf galaxy will be also supplied into the loss cone.
In this way,
gravitational interactions of the SMBH binary with these stars become effective
and the hardening rate of the SMBH binary will become high.
As a result,
the binary can close to the unstable separation
at which significant gravitational wave is emitted.

To demonstrate our scenario,
we perform highly accurate $N$-body simulations.
We show that
a merging of a dwarf galaxy is an effective process
to decrease the binding energy of a SMBH binary
and trigger the rapid shrink of the binary,
if the core of the dwarf galaxy is not destroyed
by the tidal force
and comes very close to the galactic central region.
We present our simulation model and method
in \S \ref{simulation},
numerical results in \S \ref{results},
and discussions in \S \ref{discussion}.

\section{SIMULATIONS}
\label{simulation}

\subsection{Simulation Model}

The simulation process is described in the following.
Firstly,
we make a simulation of a host galaxy with a SMBH binary
without a minor merger.
Then,
we add a dwarf galaxy
to the host galaxy
after the loss cone depletion is established,
that is,
evolution of the semi-major axis of the binary
becomes very slow.

We describe the model of a host galaxy with a SMBH binary
before a minor merger.
For the stellar distribution of the host galaxy,
we assume the King model with $W_0 =7$,
where $W_{0}$ is the nondimensional central potential of the King models.
The total mass is $M_{\rm{gal}}=1$
and the total binding energy is $E_{\rm{gal}}=-1/4$.
Here,
we use the standard $N$-body unit
in which the gravitational constant is $G=1$.
The physical unit is described in  \S \ref{unit}.
Its velocity dispersion is $\sigma _{v}=(0.5)^{1/2}$.
Two equal mass SMBHs are set
in the stellar system.
Each mass is $M_{\rm{SMBH}}=0.01$.
Initial positions and velocities of the SMBHs are
$(x,y,z)=(\pm 0.5,0,0)$ and $(v_{x},v_{y},v_{z})=(0,\pm 0.1,0)$,
respectively.
This is the same model to \citet{mak04}.

For the stellar distribution of the dwarf galaxy,
we also assume the King model.
In our all models,
the dwarf galaxies are assumed to be compact enough
to be able to come close to the galactic central region
without the destruction by the tidal force of the host galaxy.
Its mass is $M_{\rm{dwarf}}=0.1$.
Its velocity dispersion is $\sigma _{v}=(0.05)^{1/2}$.
The ratio of the velocity dispersion
of the host galaxy and the dwarf galaxy is
about $3:1$
which is expected by the cosmological simulation
\citep{kas07}.
In order to investigate effects of the compactness
and the orbit of the dwarf galaxy
on the dynamical evolution of the SMBH binary,
we assume various $W_{0}$ and various initial orbits for the dwarf galaxy.

The initial positions, the initial velocities, and $W_{0}$
of the dwarf galaxy
are shown in table~\ref{tbl-1}.
For the motion of the dwarf galaxy,
two cases are considered.
One is the zero impact parameter case
and another is the nonzero impact parameter case.
In the nonzero impact parameter cases,
the dwarf galaxy has the initial orbital angular momentum.
The specific angular momentum,
$J_{d}$,
are assumed to be $0.36$ and $0.6$
which are in the range expected
from cosmological simulations
as discussed in \S \ref{minor_merger}.
In models from $Run~1$ to $Run~3$ and in $Run~9$,
the dwarf galaxy passes through the SMBH binary directly
with the zero impact parameter.
In $Run~1$ and $Run~2$,
dwarf galaxies move in the same plane of the SMBH binary.
In $Run~3$,
it moves on the z-axis.
From $Run~4$ to $Run~8$ and in $Run~10$,
the dwarf galaxies have nonzero impact parameters initially.
In these Runs,
except for $Run~6$,
dwarf galaxies move in the prograde sense.
In $Run~6$,
it moves in a orbit tilted from the plane of the SMBH binary.
For the nondimensional central potential of the King model $W_{0}$,
$W_{0}=9$ and $W_{0}=11$ are assumed.
Such compact dwarf galaxies have been observed
in the nearby galaxies \citep{kor89}.
Their cores can be close within $r=0.2$ from the center of the host galaxy,
which is the core radius of the host galaxy,
without the destruction by the tidal force of the host galaxy.

\subsection{Physical Unit}\label{unit}

We assume that
the mass of the central region
and the velocity dispersion of the host galaxy
are $10^{10} M_{\odot}$ and $300$ km s$^{-1}$,
respectively.
Then,
the physical unit is interpreted as follows;
the unit of mass is $10^{10} M_{\odot}$,
the unit of length is about $239$ pc,
and the unit of time is about $5.51\times 10^{5}$ yr.

\subsection{Simulation Method}

We perform $N$-body simulations
of two SMBHs,
field stars in the host galaxy,
and stars in the dwarf galaxy.
From $Run~1$ to $Run~8$,
the number of $N$-body particles
is $100000$
for the stellar component in the host galaxy
and $10000$ for that in the dwarf galaxy,
respectively.
In order to investigate effects of the number of particles,
we also perform the simulation in $Run~12$ and $Run~13$
by using $200000$ and $20000$ particles
for the host galaxy and for the dwarf galaxy,
respectively.

The equations of motion for SMBHs and field stars are
\begin{equation}
\frac{d^2 \mbox{\boldmath$r$} _{BH,i}}{dt^2}
= \mbox{\boldmath$a$} _{Bf,i}
+ \mbox{\boldmath$a$} _{BB,i}
\end{equation}
\noindent
and
\begin{equation}
\frac{d^2 \mbox{\boldmath$r$} _{f,i}}{dt^2}
= \mbox{\boldmath$a$} _{ff,i}
+ \mbox{\boldmath$a$} _{fB,i},
\end{equation}
\noindent
respectively,
where $\mbox{\boldmath$a$} _{Bf,i}$ is the acceleration
on the SMBH from field stars,
$\mbox{\boldmath$a$} _{BB,i}$ is the acceleration
on the SMBH from another SMBH,
$\mbox{\boldmath$a$} _{ff,i}$ is the acceleration
on the field star from other field stars,
and $\mbox{\boldmath$a$} _{fB,i}$ is the acceleration
on the field star from SMBHs.
The softening lengths
between field stars,
SMBHs and field stars,
and SMBHs
are $\epsilon _{\rm{ff}}=10^{-4}$,
$\epsilon _{\rm{fB}}=10^{-6}$,
and $\epsilon _{\rm{BB}}=10^{-6}$,
respectively,
in order to resolve much less than a sub-pc scale.
The effect of gravitational wave is not considered.

The fourth-order Hermite scheme \citep{mak92}
is used for time integration.
The predictors are

\begin{equation}
\mbox{\boldmath$x$}_p(t_0+\Delta t) =
\mbox{\boldmath$x$}_0(t_0)
+\mbox{\boldmath$v$}_0(t_0) \Delta t
+\frac{1}{2}\mbox{\boldmath$a$}_0(t_0) \Delta t^2
+\frac{1}{6}\dot{\mbox{\boldmath$a$}}_0(t_0) \Delta t^3
\end{equation}

\noindent
and

\begin{equation}
\mbox{\boldmath$v$}_p(t_0+\Delta t) =
\mbox{\boldmath$v$}_0(t_0)
+\mbox{\boldmath$a$}_0(t_0) \Delta t
+\frac{1}{2}\dot{\mbox{\boldmath$a$}}_0(t_0) \Delta t^2.
\end{equation}

\noindent
The correctors are

\begin{equation}
\mbox{\boldmath$x$}_c(t_0+\Delta t) =
\mbox{\boldmath$x$}_p
+\frac{1}{24}\mbox{\boldmath$a$}^{(2)} \Delta t^4
+\frac{1}{120}\mbox{\boldmath$a$}^{(3)} \Delta t^5
\end{equation}

\noindent
and 

\begin{equation}
\mbox{\boldmath$v$}_c(t_0+\Delta t) =
\mbox{\boldmath$v$}_p
+\frac{1}{6}\mbox{\boldmath$a$}^{(2)} \Delta t^3
+\frac{1}{24}\mbox{\boldmath$a$}^{(3)} \Delta t^4,
\end{equation}

\noindent
$\mbox{\boldmath$a$}^{(2)}$
and $\mbox{\boldmath$a$}^{(3)}$ are

\begin{equation}
\mbox{\boldmath$a$}^{(2)}
= \frac{-6(\mbox{\boldmath$a$}_0-\mbox{\boldmath$a$}_1)
-\Delta t
(4\dot{\mbox{\boldmath$a$}}_0
+2\dot{\mbox{\boldmath$a$}}_1)}
{\Delta t^2}
\end{equation}

\begin{equation}
\mbox{\boldmath$a$}^{(3)}
= \frac{12(\mbox{\boldmath$a$}_0-\mbox{\boldmath$a$}_1)
+6\Delta t
(\dot{\mbox{\boldmath$a$}}_0
+\dot{\mbox{\boldmath$a$}}_1)}
{\Delta t^3},
\end{equation}

\noindent
where $\mbox{\boldmath$a$}_1$
and $\dot{\mbox{\boldmath$a$}}_1$
is the acceleration and its time derivative at $t=(t_{0}+\Delta t)$.
The individual timesteps are combined to this scheme
\citep{mak91}.
The timestep formula is given by

\begin{equation}
 \Delta t_{i}
 =\sqrt{\eta \frac{|\mbox{\boldmath$a$}_{i}|
 |\mbox{\boldmath$a^{(2)}$}_{i}|
 +|\mbox{\boldmath$\dot{a}$}_{i}|^{2}}
 {|\mbox{\boldmath$\dot{a}$}_{i}|
 |\mbox{\boldmath$a^{(3)}$}_{i}|^2
 +|\mbox{\boldmath$a^{(2)}$}_{i}|^{2}}},
\end{equation}

\noindent
where $\eta$ is the parameter
which controls the integration accuracy.
In our simulations,
we adopt $\eta = 0.005$ for SMBHs
and $\eta = 0.02$ for field stars,
respectively.
The acceleration and the time derivative of the acceleration
by field stars
are calculated
by GRAPE-6 \citep{mak03}
which is a special purpose hardware
to make those calculations very fast.
Those by SMBHs are calculated
by a host computer
in order to make energy error small.
In our all simulations,
the error of the total energy is less than
$0.1$ \% of the initial total energy.

\section{RESULTS}
\label{results}

\subsection{Loss Cone Depletion}

\begin{figure}
\begin{center}
 \FigureFile(40mm,40mm){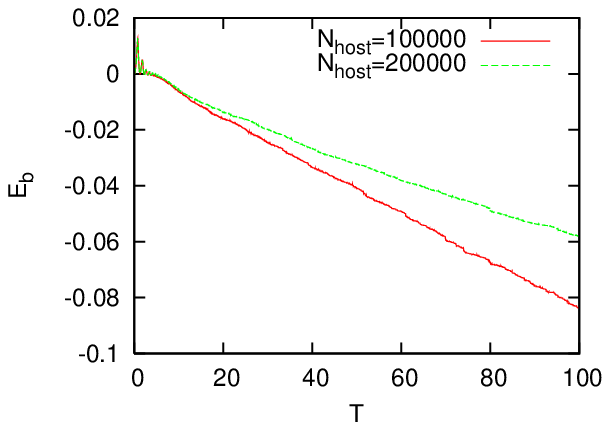}
 \FigureFile(40mm,40mm){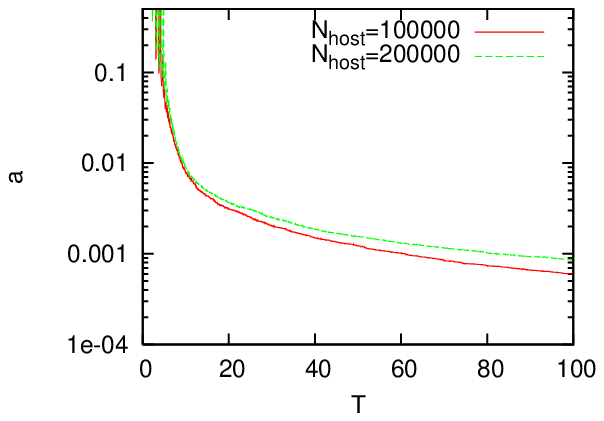}
\end{center}
\caption{The time evolution of
the binding energy (left)
and the semi-major axis (right)
of the SMBH binary
in the host galaxy.
The red and blue lines are
for the cases of
$N_{\rm{host}}=100000$ and $N_{\rm{host}}=200000$,
respectively,
where $N_{\rm{host}}$ is the particle number of the host galaxy.}
\label{host}
\end{figure}

Figure~\ref{host} shows the time evolution
of the binding energy
and the semi-major axis of the SMBH binary
without a minor merger.
The SMBH binary loses its binding energy
because of the dynamical friction from field stars.
As a result,
the semi-major axis shrinks rapidly.
After $T=20$,
the hardening rate,
which is defined by
$\beta \equiv |\Delta E _{b} / \Delta t|$,
becomes very low
and is almost constant.
They are about $\beta = 0.0008$ and $\beta = 0.0006$
in the simulations with $N_{\rm host}=100000$
and $N_{\rm host}=200000$,
respectively.
Then,
the semi-major axis hardly shrinks.

\begin{figure}
\begin{center}
 \FigureFile(40mm,40mm){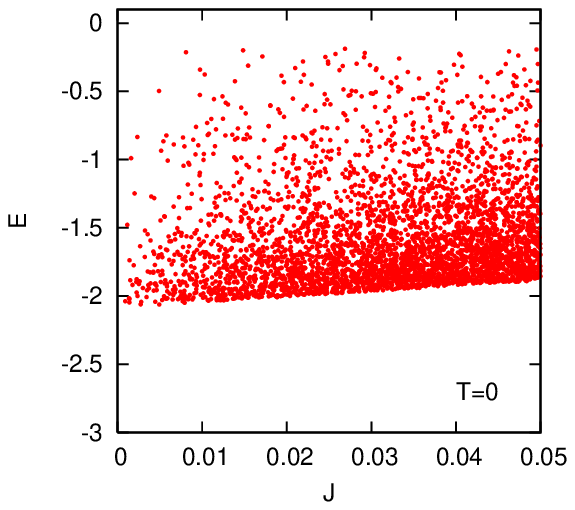}
 \FigureFile(40mm,40mm){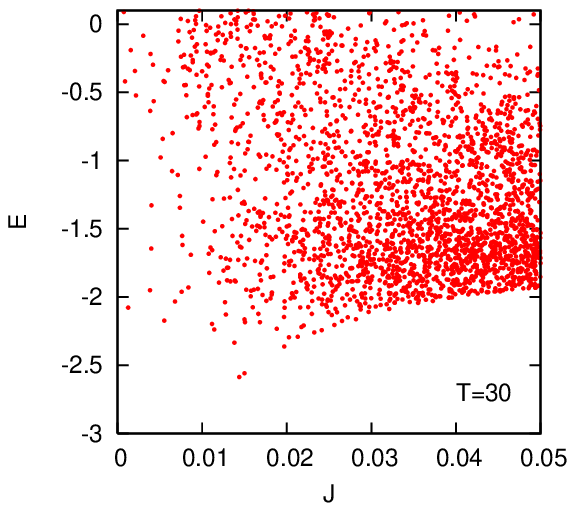}
\end{center}
\caption{
Distributions of the field stars in the host galaxy
of $N_{\rm{host}}=100000$
in the $(J,E)$ plane at $T=0$ (left)
and at $T=30$ (right)
at which the loss cone depletion is already established.
}
\label{JE_lc}
\end{figure}

The reason that the hardening rate becomes low is
the loss cone depletion.
As evidence of the loss cone depletion,
in Fig.~\ref{JE_lc} we show the distribution of star particles
at $T=0$ and $T=30$
in $(J,E)$ plane
where $J$ is a specific angular momentum
about the center of mass of the galaxy
and $E$ is a specific binding energy
of each star particle.
The number of the star particles
with low $J$ and $E$,
which are supplied into the loss cone,
decreases significantly
from the initial state
at $T=0$ to $T=30$
at which the hardening rate is already low.
This is clear evidence of the loss cone depletion.

The hardening rate and time evolution of the semi-major axis of the binary
strongly depend on the particle number of the host galaxy.
The hardening rate is smaller in higher resolution simulations
with a larger number of $N$-body particles.
This is because
the timescale of the two body relaxation
by which star particles are supplied into the loss cone
is longer in the simulation with a larger number of particles.
This property is reported
by \citet{mak04}.

\subsection{Effects of a Minor Merger}

We add the dwarf galaxy to the host galaxy
at $T=30$
when the loss cone depletion is already realized.

\subsubsection{Minor mergers of zero impact parameter}

\begin{figure}
\begin{center}
 \FigureFile(40mm,40mm){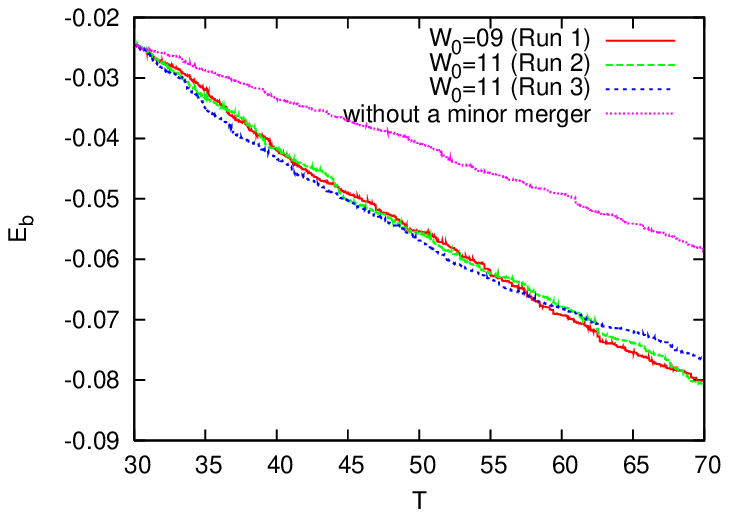}
 \FigureFile(40mm,40mm){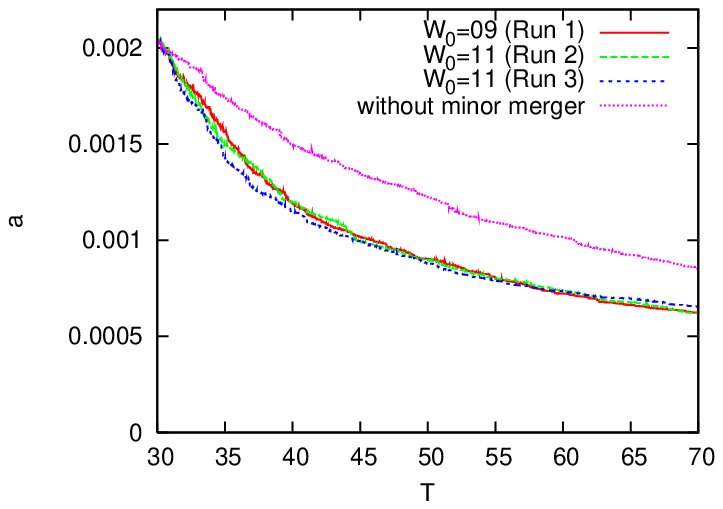}
\end{center}
\caption{
The time evolution of
the binding energy (left)
and the semi-major axis of the SMBH binary (right)
in the cases
of the zero impact parameter
($Run~1$-$3$).
For comparison,
the result without a minor merger is also shown by the light blue line.
}
\label{a_0impact}
\end{figure}

In Fig.~\ref{a_0impact},
we show the time evolution
of the binding energy and the semi-major axis of the SMBH binary
from $Run~1$ to $Run~3$
in which the dwarf galaxy moves with zero impact parameter.
After the dwarf galaxy passes through the center of the host galaxy
at $T=31$,
the hardening rate becomes high.
The average hardening rate from $T=31$ to $T=60$
is $\beta=0.0015-0.0016$ in all these models,
which is about 2.0 times higher than
that of the case without a minor merger
$\beta=0.0008$.
Although the core of the dwarf galaxy is destroyed
by the tidal force of the SMBHs
after the first encounter with the binary
in these simulations,
the high hardening rate continues.
In these cases,
the minor merger reduces the binding energy of the binary
effectively.
As a result,
the rapid orbital decay of the binary occurs
and the semi-major axis shrinks rapidly.
The rate of the shrink becomes similar to the case without a minor merger
after about $T=45$.

The high hardening rate and the rapid shrink of the semi-major axis
are caused by following process:
The core of the dwarf galaxy passes through
the central region of the host galaxy
without the destruction by the tidal force,
since the density of the core is higher than that of the host galaxy.
When the core is close to the center
of the host galaxy,
it perturbs the gravitational potential of the host galaxy.
Due to the perturbed potential,
orbits of star particles change
and then a large number of star particle orbits are able to pass through
the loss cone.

\begin{figure}
\begin{center}
 \FigureFile(40mm,40mm){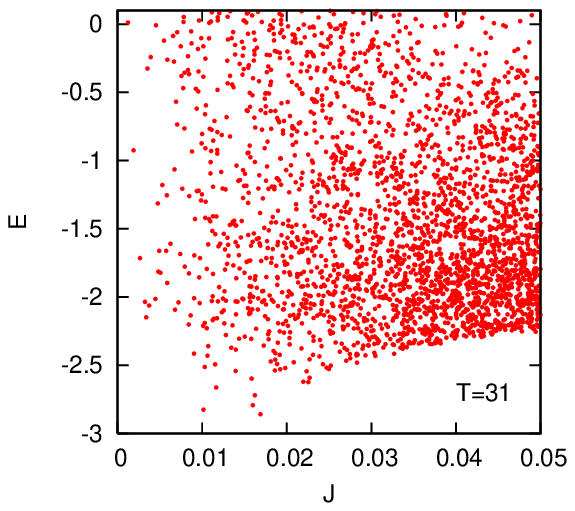}
 \FigureFile(40mm,40mm){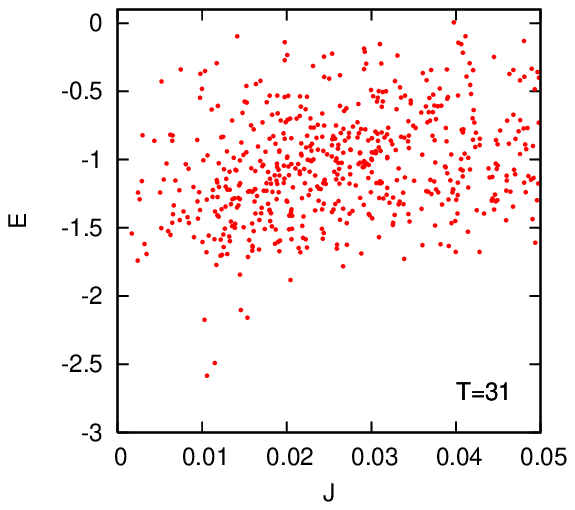}
\end{center}
\caption{
Distributions of the field stars
in the host galaxy (left)
and the dwarf galaxy (right)
in the $(J,E)$ plane
at $T=31$
in $Run 2$.
}
\label{JET31Run4}
\end{figure}

In Fig.~\ref{JET31Run4},
we show the distribution of the star particles of the host galaxy (left panel)
and the dwarf galaxy (right panel)
in the $(J,E)$ plane in $Run~2$
at $T=31$,
here the center is set at the center of gravity of the binary.
In the left panel,
the star particles
with low $J$ and low $E$
increase in comparison with
the right panel of Fig.~\ref{JE_lc}
in which the loss cone depletion is established.
These star particles of the host galaxy are supplied into the loss cone.
In the right panel of Fig.~\ref{JET31Run4},
there are star particles with low $J$.
Such star particles of the dwarf galaxy are also supplied into the loss cone.
These star particles are able to interact gravitationally with SMBHs.

In the case of the zero impact parameter,
the hardening rate of the SMBH binary becomes high
and the semi-major axis shrinks rapidly
in all dwarf galaxy models
with $W_{0}=9$ and $W_{0}=11$.
Those evolution does not depend on
the compactness of the dwarf galaxy,
$W_{0}$.
The numerical results show that
the minor merger
during which the dwarf galaxy passes through the binary
is an effective mechanism
to decrease the binding energy of the binary.

\subsubsection{Minor mergers of nonzero impact parameter}

\begin{figure}
\begin{center}
 \FigureFile(40mm,40mm){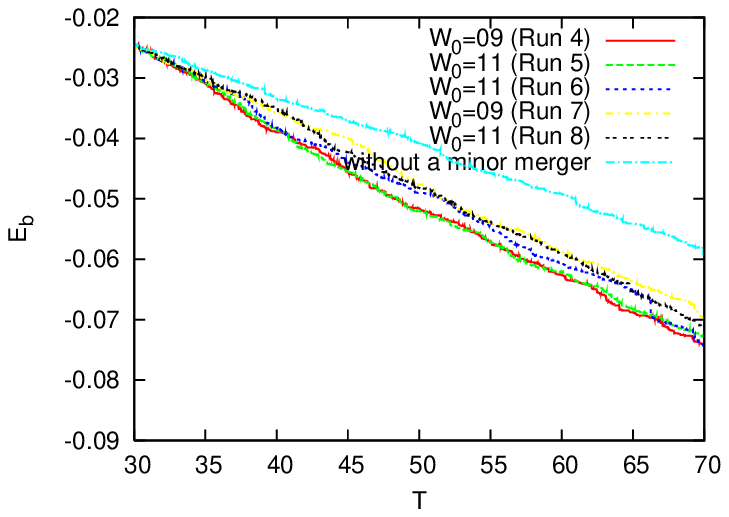}
 \FigureFile(40mm,40mm){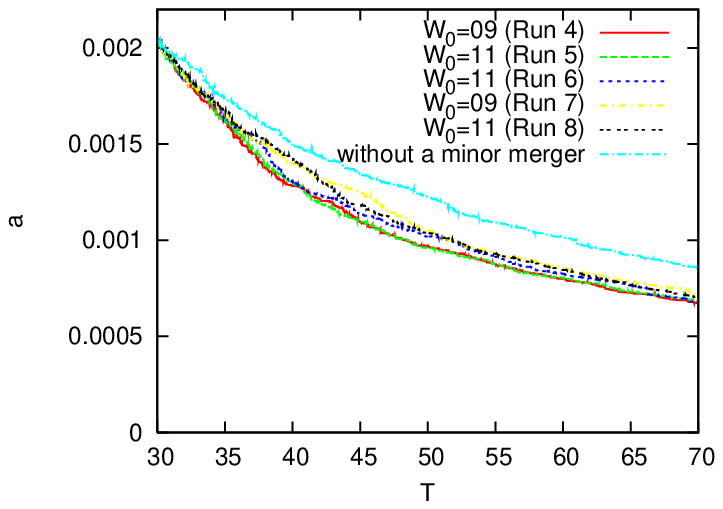}
\end{center}
\caption{
The time evolution of
the binding energy (left)
and the semi-major axis of the SMBH binary (right)
in the cases
of the nonzero impact parameter
($Run~4$-$8$).
For comparison,
the result without a minor merger is also shown by the light blue line.
}
\label{Ea_impact}
\end{figure}

We show the results of the dwarf galaxy model
with the nonzero impact parameter.
The time evolution of
the binding energy and the semi-major axis
are shown in Fig.~\ref{Ea_impact}.

In $Run~4$, $Run~5$, and $Run~6$
in which initial specific angular momentum of the dwarf galaxy is $J_{d}=0.36$,
the time evolution of the binding energy is
resemble to the results of the dwarf galaxy models
with the zero impact parameter.
The hardening rate becomes high
at $T\sim 31$
when the dwarf galaxy comes close to
the galactic central region.
The hardening rate is about $\beta=0.0013-0.0014$.
It continues to the end time of our simulations.
The rate is much higher than the case without a minor merger
and little lower than that of the dwarf galaxy models
with the zero impact parameter.

\begin{figure}
\begin{center}
 \FigureFile(40mm,40mm){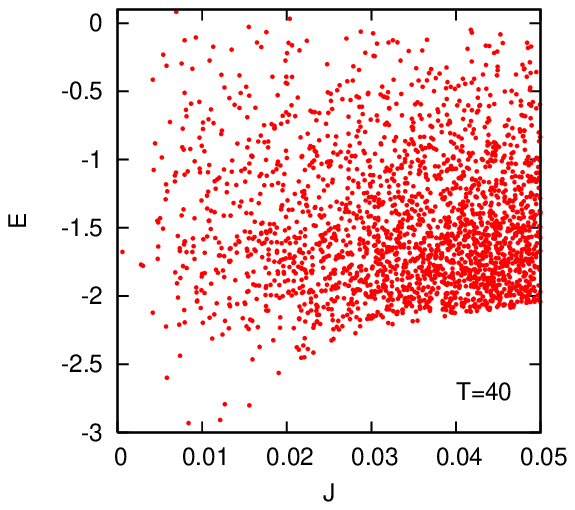}
 \FigureFile(40mm,40mm){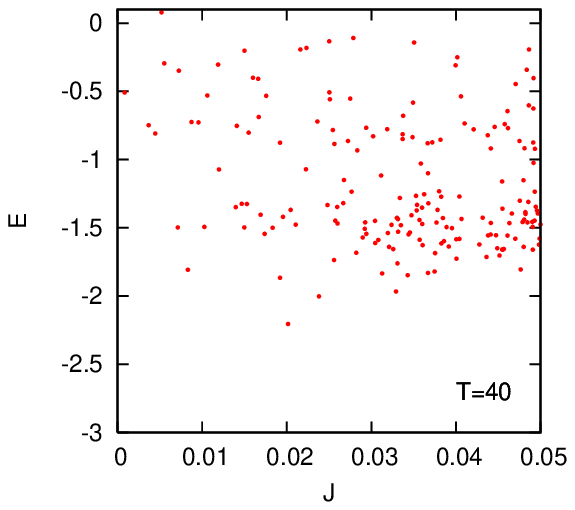}
\end{center}
\caption{
Distributions of the field stars
in the host galaxy (left)
and the dwarf galaxy (right)
in the $(J,E)$ plane
at $T=40$
in $Run 5$.
}
\label{JET40Run7}
\end{figure}

The distribution of the star particles in the $(J,E)$ plane
at $T=40$ in $Run~5$
is shown in Fig.~\ref{JET40Run7}.
In the left panel
which shows the distribution of the star particles of the host galaxy,
the number of star particles with low $J$ and low $E$ increase
in comparison with the right panel of Fig.~\ref{JE_lc}.
The number of such star particles is
almost similar to the cases
of the dwarf galaxy models with the zero impact parameter.
These star particles are supplied into the loss cone.
However,
there are a few star particles of the dwarf galaxy
with low $J$ and low $E$
in the right panel
which shows the distribution of the star particles of the dwarf galaxy.
This result indicates that
star particles of the dwarf galaxy are hard to be supplied into the loss cone,
contrary to the cases of the zero impact parameter.
In these cases,
the SMBH binary loses the binding energy
mainly by disturbed stars of the host galaxy.
Since there are a few star particles of the dwarf galaxy
which interact with the binary,
the hardening rate is slightly lower than
the dwarf galaxy models with the zero impact parameter.

Since the SMBH binary loses the binding energy effectively,
the semi-major axis shrinks rapidly.
The rapid shrink occurs
as soon as the dwarf galaxy comes close to the galactic central region.
It continues until about $T=45$.
After about $T=45$,
the decrease rate of the semi-major axis
becomes same as the case without a minor merger.

For $J_{d}=0.36$,
time evolution of the binding energy and of the semi-major axis
does not depend on the compactness of the dwarf galaxies,
$W_{0}$,
although the destruction timescale
of the core of the dwarf galaxy for each model
is different,
for examples,
the core is destroyed at about $T=35$ in $Run~4$
and at about $T=45$ in $Run~5$.

In $Run~7$ and $Run~8$,
in which initial specific angular momentum of the dwarf galaxy is $J_{d}=0.6$,
increase of the hardening rate of the binary is delayed
till the dwarf galaxy comes close to the galactic central region.
After the core is close to the galactic central region within about $r=0.2$
without their destruction by the tidal force,
the hardening rate and time evolution of the semi-major axis
become high
similarly to those in $Run~4$-$6$.

These results show that
the dwarf galaxy with nonzero impact parameter
also increases the hardening rate of the SMBH binary,
since it can disturb orbits of star particle by its gravitational potential
and such star particles are supplied into the loss cone.

\subsubsection{Effects of the particle number}

\begin{figure}
\begin{center}
 \FigureFile(40mm,40mm){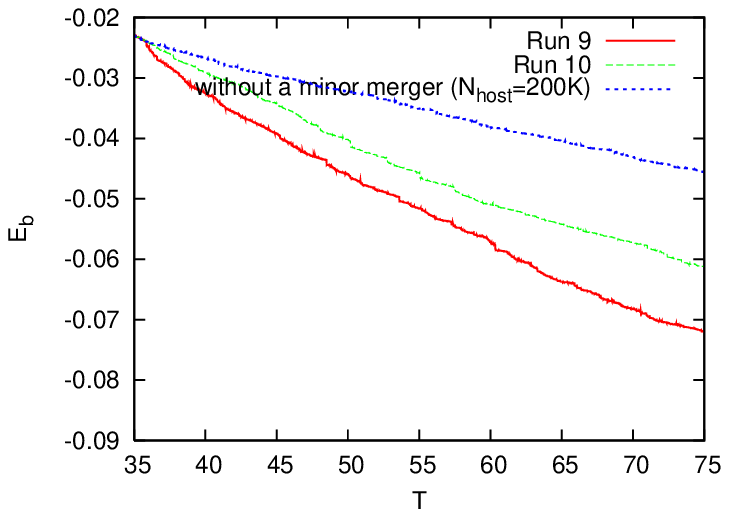}
 \FigureFile(40mm,40mm){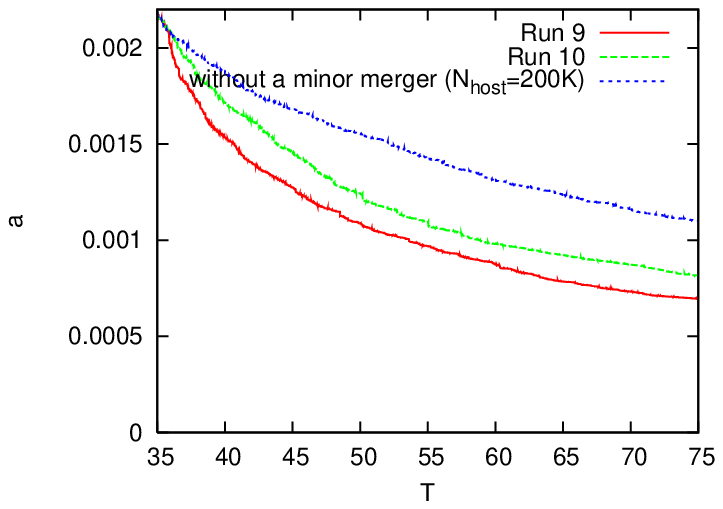}
\end{center}
\caption{
The time evolution of
the binding energy (left)
the semi-major axis of the SMBH binary (right)
in the case
that particle numbers are $N_{\rm{host}}=200000$
and $N_{\rm{dwarf}}=20000$.
For comparison,
the result without a minor merger is also shown by the blue line.
}
\label{N200K}
\end{figure}

We perform simulations
with $N_{\rm{host}}=200000$ and $N_{\rm{dwarf}}=20000$
in order to investigate effects of the particle number.
The dwarf galaxy is added to the host galaxy at $T=35$
when the semi-major axis of the SMBH binary
is similar scale to that at $T=30$
in the simulation with $N_{\rm{host}}=100000$.
The dwarf galaxy models correspond to $Run 2$ and $Run 5$.
The time evolution of
the binding energy and the semi-major axis of the binary
is shown in Fig.~\ref{N200K}.

Time evolution of the binding energy and semi-major axis of the binary
is similar to the results of $N_{\rm{host}}=100000$.
The hardening rate becomes high
after the dwarf galaxy comes close to
the galactic central region.
The average hardening rates from $T=35$ to $T=60$
are $\beta = 0.0014-0.0015$
in $Run~9$
and $\beta = 0.0012-0.0013$
in $Run~10$,
which is much larger than the case without a minor merger.
As a result,
the semi-major axis shrinks rapidly.
This result confirms that
a minor merger triggers rapid shrink
of a SMBH binary
in higher resolution simulations.

The effect of a minor merger
on time evolution of the hardening rate and the semi-major axis
becomes clear
in the simulation of $N_{\rm{host}}=200000$
than the results of $N_{\rm{host}}=100000$.
This is because
the timescale of the two body relaxation of $N$-body particles
is longer in the simulation of $N_{\rm{host}}=200000$
and the number of supplied star particles by the two body relaxation
is less than
the simulation of $N_{\rm{host}}=100000$.
Therefore,
effects of the two body relaxation become fewer
and effects of a minor merger become clear.
This result indicates that
the effects of minor mergers appear clearly
in higher resolution simulations.

\section{DISCUSSION}
\label{discussion}

\subsection{Minor Mergers of the Compact Dwarf Galaxies}
\label{minor_merger}

We demonstrate
that separation of a SMBH binary shrinks rapidly
after a compact dwarf galaxy comes close to
the central region of a host galaxy.
It is important for our scenario
that the dwarf galaxies are compact
and are able to come close to the galactic central region
without their destruction by the tidal force of the host galaxy.
In this section,
we discuss the possibility that such minor mergers occur.

Dwarf galaxies formed in the early universe are compact,
since the mean density of dark matter halos is higher
in the earlier universe,
$\rho _{\rm{DM}}(z)\propto (1+z)^{-3}$.
Therefore,
many compact dwarf galaxies
are expected to form in the early universe
and merge to their host galaxy.
This is confirmed
by the cosmological numerical simulations
of the galaxy formation (e.g., \citet{sai06}).

To trigger the rapid shrink of a SMBH binary,
such compact dwarf galaxies need to come close to
the central region of a host galaxy.
To investigate the possibility,
we calculate the motions of the dwarf galaxies
with various initial orbital parameters
which are in the range expected
from the cosmological numerical simulations.
Here,
the fourth order Runge-Kutta method is used for the time integration.
We assume that
the dark halo and stellar potentials of the host galaxy are fixed.
For the dynamical friction force,
we use the formula given by \citet{fuk92}.
The initial position of the dwarf galaxy is set at $50$ kpc
from the center of the host galaxy
which is vicinity of the virial radius of the dark halo.
The initial parameter of
the nondimensional orbital angular momentum
of the dwarf galaxy
is from $\lambda=0.01$ to $0.04$
which is the range of the spin parameter distribution
of sub halos in a host dark halo \citep{sha05}.

\begin{figure}
\begin{center}
 \FigureFile(40mm,40mm){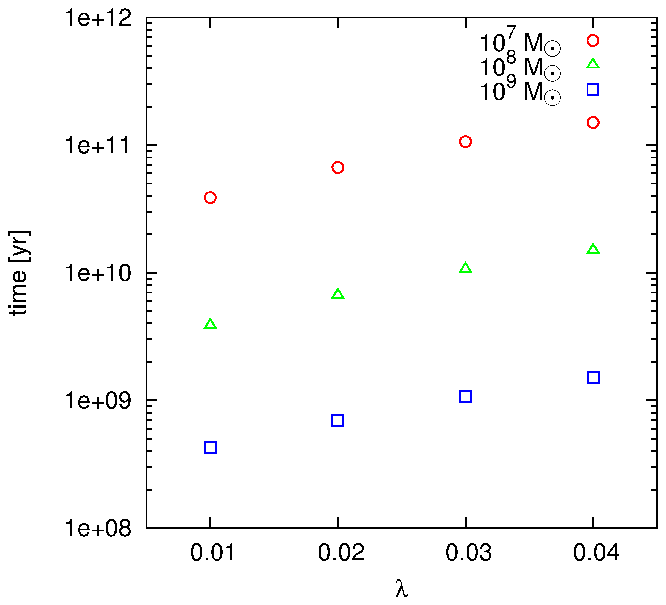}
 \FigureFile(40mm,40mm){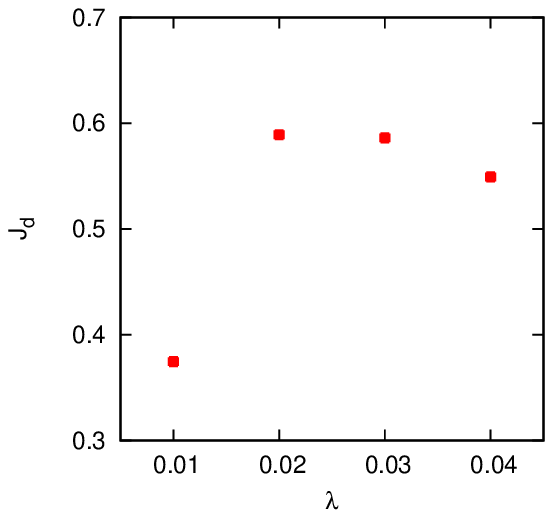}
\end{center}
\caption{
The time that a dwarf galaxy sinks
from $50$ kpc to $100$ pc from the galactic center (left).
The horizontal axis shows the spin parameter,
$\lambda$.
The circles, triangles, and  squares denote the cases that
the masses of the dwarf galaxies are
$M_{\rm{dwarf}}=10^7 M_{\odot}$, $10^8 M_{\odot}$,
and $10^9 M_{\odot}$,
respectively.
The angular momentum,
$J_{d}$,
of the dwarf galaxy of $M_{\rm{dwarf}}=10^9 M_{\odot}$
at the time
when the dwarf galaxies
with the initial spin parameter from $\lambda=0.01$ to $\lambda=0.04$
pass through
$r=1$ from the center of the host galaxy
in our unit (right).
}
\label{lambda_dwarf}
\end{figure}

The time
which is needed for the dwarf galaxies to move to the galactic central region
(within $100$ pc)
is shown in the left panel of Fig.~\ref{lambda_dwarf}.
The dwarf galaxy with $10^9 M_{\odot}$
can move to the galactic center
for the spin parameter $\lambda =0.01$-$0.04$
within $2\times 10^{9}$ yr.
Such dwarf galaxies can come close to the galactic center
within much less than $10^{10}$ yr.
For the dwarf galaxies with $10^{8} M_{\odot}$,
they can come close to the center
within $10^{10}$ yr
in the case of the spin parameter
of $\lambda = 0.01$ and $\lambda = 0.02$.

The right panel of Fig.~\ref{lambda_dwarf} shows
the specific angular momentum of the dwarf galaxy
when it passes through $r=1$ from the galactic center of the host galaxy.
In the case of $\lambda = 0.01$,
the specific angular momentum is about $0.37$.
For larger $\lambda$,
it ranges from $0.5$ to $0.6$.
The models from $Run~4$ to $Run~6$ and $Run~10$
correspond to the case of $\lambda = 0.01$.
The models of $Run~7$ and $Run~8$
correspond to the case of $\lambda = 0.02-0.04$.
Then,
it is needed for our scenario
that dwarf galaxies have mass more than $10^8 M_{\odot}$
and smaller $\lambda$ than $\lambda=0.03$.

\subsection{Conclusions}

We have performed $N$-body simulations
and have shown that
a minor merger is an effective process
to resolve the loss cone depletion problem.
If the core of the dwarf galaxy is not destroyed by the tidal force
and comes close to the galactic central region,
disturbed stars of the host galaxy are supplied into the loss cone.
If the dwarf galaxy passes through the SMBH binary directly,
stars of the dwarf galaxy are also supplied into the loss cone.
After that,
SMBHs can interact gravitationally with these stars
and the binary loses its binding energy.
In this process,
three body interactions of the SMBH binary with these stars should be important
\citep{per07,per08}.
As a result,
the hardening rate of the binary becomes high
and the semi-major axis can shrink rapidly
in the host galaxy.

We have also performed high resolution simulations
of $200000$ $N$-body particles
with for the host galaxy.
We have confirmed that
the minor merger triggers
the high hardening rate and the rapid shrink
of the SMBH binary
in the high resolution simulations.
We find that
the difference of time evolution of the hardening rate and the semi-major axis
between with and without a minor merger
becomes clear
in higher resolution simulations.
This is because
timescale of the two body relaxation
of $N$-body particles
becomes longer
in the $200000$ $N$-dody simulations.
This result indicates that
effects of the minor merger appear clearly
in the realistic stellar system
in which the two body relaxation time is larger than the Hubble time.

It is important for our scenario that
the dwarf galaxy is enough compact
to come close to the galactic central region.
In the hierarchical galaxy formation scenario,
such minor mergers are expected to occur frequently.
Therefore,
we emphasize that
our scenario is one of the effective processes
to trigger the rapid orbital decay of the SMBH binary
and helps its coalescence within the Hubble time
together with other previous proposed mechanisms.

\bigskip
The authors appreciate the anonymous referee
for helpful advices.
We thank Junichro Makino
for helpful advices and suggestions.
We also thank Keiichi Wada, Koji Tomisaka, Takayuki R. Saitoh,
Masayuki Y. Fujimoto, and Kazuo Sorai
for their helpful discussions,
and Masaki Iwasawa and Tamon Suwa
for their helpful advices on simulation method.
This work has been supported by
Grant-in-Aid for the 21st Century COE Scientific Research Programme
on ``Topological Science and Technology''
from the Ministry of Education, Culture, Sport, Science,
and Technology of Japan (MECSST),
in part by Grant-in-Aid for Scientific Research
(14340058)
of Japan Society for the Promotion of Science,
and in part by Hokkaido University Grant Program
for New Fusion of Extensive Research Fields.
Numerical computations were carried out
on GRAPE system
at Center for Computational Astrophysics,
CfCA,
of National Astronomical Observatory of Japan.

\begin{table}
 \caption{Dwarf galaxy models
and the particle number of the host galaxy and the dwarf galaxy.}
\label{tbl-1}
 \begin{tabular}{lccccc}
 \hline
Run & $(x,y,z)$ & $(v_{x},v_{y},v_{z})$ &  $W_{0}$ &
$J_d$ &
$N_{\rm{host}+\rm{dwarf}}$ \\
 \hline
1 & $(0,-1,0)$ & $(0.0,0.7,0.0)$ &  9  & 0.0 & $110000$ \\
2 & $(0,-1,0)$ & $(0.0,0.7,0.0)$ & 11 & 0.0 & $110000$ \\
3 & $(0,0,-1)$ & $(0,0.0,0.7)$ & 11 & 0.0 & $110000$ \\
4 & $(0,-1,0)$ & $(0.36,0.6,0.0)$ & 9 & 0.36 & $110000$ \\
5 & $(0,-1,0)$ & $(0.36,0.6,0.0)$ & 11 & 0.36 & $110000$ \\
6 & $(0,-1,0)$ & $(0.0,0.6,0.36)$ & 11 & 0.36 & $110000$ \\
7 & $(0,-1,0)$ & $(0.6,0.36,0.0)$ & 9 & 0.6 & $110000$ \\
8 & $(0,-1,0)$ & $(0.6,0.36,0.0)$ & 11 & 0.6 & $110000$ \\
9 & $(0,-1,0)$ & $(0.0,0.7,0.0)$ & 11 & 0.0 & $220000$ \\
10 & $(0,-1,0)$ & $(0.36,0.6,0.0)$ & 11 & 0.36 & $220000$ \\
 \hline
\end{tabular}
\end{table}

\end{document}